# MEMRISTOR EQUATIONS: INCOMPLETE PHYSICS AND UNDEFINED PASSIVITY/ACTIVITY

KYLE SUNDQVIST [1], DAVID K. FERRY [2] and LASZLO B. KISH [3†]

[1] *Department of Physics, San Diego State University, 5500 Campanile Drive, San Diego, CA 92182-1233 USA*

[2] *School of Electrical, Computer, and Energy Engineering and Center for Solid State Electronic Research, Arizona State University, Tempe, AZ 85287-5706, USA*

[3] *Department of Electrical and Computer Engineering, Texas A&M University, TAMUS 3128, College Station, TX 77843-3128*

**Abstract.** In his seminal paper, Chua presented a fundamental physical claim by introducing the memristor, "The missing circuit element". The memristor equations were originally supposed to represent a passive circuit element because, with active circuitry, arbitrary elements can be realized without limitations. Therefore, if the memristor equations do not guarantee that the circuit element can be realized by a passive system, the fundamental physics claim about the memristor as "missing circuit element" loses all its weight. The question of passivity/activity belongs to physics thus we incorporate thermodynamics into the study of this problem. We show that the memristor equations are physically incomplete regarding the problem of passivity/activity. As a consequence, the claim that the present memristor functions describe a passive device lead to unphysical results, such as violating the Second Law of thermodynamics, in infinitely large number of cases. The seminal memristor equations cannot introduce a new physical circuit element without making the model more physical such as providing the Fluctuation Dissipation Theory of memristors.

## 1. Introduction

*1.1 On Chua's memristor model*

In 1971, Chua introduced the memristor, a new mathematical circuit theory element [1] that, to be fundamentally interesting, was supposed to be a passive device. The mathematical model interrelates the time-integral of the voltage (voltage flux) on the memristor

$$\Phi = \int_0^t U(\tau) d\tau \qquad (1)$$

with the time-integral of the current (charge) through the device

$$q = \int_0^t I(\tau) d\tau \qquad (2)$$

as follows:

$$M(q) = \frac{d\Phi}{dq}, \qquad (3)$$

$$U(t) = M[q(t)]I(t) \qquad (4)$$

---
[†] Corresponding Author

where $M(q)$ is the memristor function. It is claimed that for the memristor to be a *passive* circuit element the following condition must be satisfied:

$$M(q) \geq 0 \qquad (5)$$

(Note, the lower boundary of the integral is often taken from minus infinity, which is unphysical. The more physical time coordinate "zero" can be the moment of time when the memristor comes into existence, and an initial value of the integral can be given, if needed. However, this issue has no importance in our present paper).

If the memristor function is constant, $M_0$, then the memristor and Equation 4 represents a resistor of resistance $R_0$, that is, $M(q) = M_0$, where $R_0 = M_0$. Then Equation 4 becomes

$$U(t) = R_0 I(t) \ . \qquad (6)$$

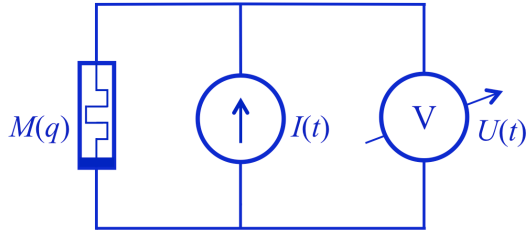

**Figure 1.** Memristor driven by a current generator while the voltage on it is monitored. For definitions, see Equations 1-5 in the text.

One of the most fundamental questions about Chua's memristor model is:

*Do Equations 1-5 indeed always represent a passive device, or there are situations where the realization of a memristor function requires an active device?*

With active circuitry, arbitrary elements can be realized without limitations. Therefore, *if the memristor equations do not guarantee that the circuit element can be realized by a passive system, the fundamental physics claim about the memristor as "the missing circuit element" loses all its weight*.

It is important to note that many nonlinear impedances with memory effects exist since the existence of electronics that are passive devices and can be called memristors (for recent ones, see, e.g. [2,3]). These devices are not of our interest because they are examples for practical passive devices and they do not answer our fundamental physical question above about the generality of the memristor claim.

The question of passivity/activity is a physical problem and former engineering definitions of this matter were shown self-contradictory and unphysical at certain practical conditions where they lead to perpetual motion machines [4]. Therefore, we will use the most advanced, statistical-thermodynamic definition of passivity/activity based on statistical physics that works correctly even with thermal noise (the quoted text below is from [4]):

"a) The device-in-question is active if the following condition holds. Suppose we have a hypothetical-device, which does not require an external energy source to function and has the same signal-response characteristics as the device-in-question. In an isolated system with thermal equilibrium, such a hypothetical-device in a proper circuit would be able to produce steady-state entropy reduction in the system that is originally in thermal equilibrium, where the other elements are all passive. In other words, such hypothetical device would violate the Second Law of Thermodynamics.

Note: for example, such an entropy reduction could be a steady-state positive or negative temperature gradient in the device's environment; etc.; anything that can break thermal equilibrium conditions and persists over the duration of operation of the device.

b) Physical implication: Such a hypothetical-device cannot exist in practice, thus an active physical device always requires an external energy source to execute its response characteristics of activity.

c) The device-in-question is passive if it is not active."

The structure of the paper is as follows:

In Section 2, we briefly outline the Fluctuation Dissipation Theorem of classical statistical physics.

In Section 3, we show why the above memristor equations are only abstract circuit theoretical models that are unphysical and/or incomplete in their present form and that the present model requires an active device in an infinite number of situations including the case when the memristor emulates a resistor.

## 2. Fluctuation-Dissipation Theorem and thermal noise of impedances and resistors

Contrary to Chua's memristor model (Equations 1-6), the statistical thermodynamics of the three fundamental circuit elements, that is, the resistor, capacitor and inductor, is sufficiently defined via the Fluctuation-Dissipation Theorem [5-8]. It states that, in the classical physical limit, any impedance combined from these elements will act as a thermal noise generator with an internal impedance value identical to this particular impedance.

The Second Law of Thermodynamics requires [5-8] that, in thermal equilibrium, the time-average of the instantaneous power flow between two parallel impedances is zero, *i.e.*,

$$\langle P_{a \to b}(t,T) \rangle_t = 0 \quad , \qquad (7)$$

where $t$ is time and $P_{a \leftrightarrow b}(t,T)$ denotes instantaneous power flow between resistors $Z_a$ and $Z_b$, as illustrated in Fig. 2. This expression holds for any frequency range.

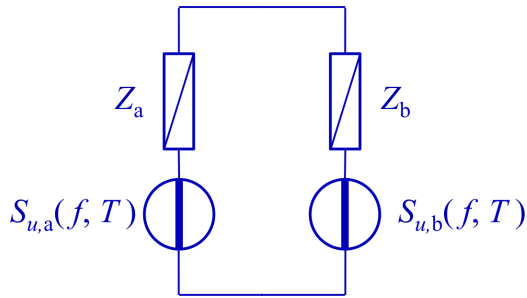

**Figure 2**. Circuit for discussing Johnson noise and representing net power flow between resistors $Z_a$ and $Z_b$. The voltage generators produce $S_u(f,T)$ power density spectra at frequency $f$ and temperature $T$.

For a *passive* impedance $Z(f)$ in thermal equilibrium, Equation 7 demands that the functional form of the power density spectral function of noise voltage $U(t)$ is [8]:

$$S_u(f,T) = \mathrm{Re}[Z(f)] Q(f,T) = R(f) Q(f,T) \quad , \qquad (8)$$

where $R(f)$ is the real part of the impedance and $S_u(f)$ represents the power density spectrum of thermal (Johnson) noise voltage. $Q(f,T)$ is a universal function of frequency and temperature and it is independent of material properties, geometry, and mechanism for electrical conduction [5-8].

In the low-frequency (classical physical limit):

$$Q(f,T) = 4kT \quad . \tag{9}$$

For active impedances, there is no related fundamental law and the noise is determined by also other factors not visible in Equation 8.

For inductors and capacitors, Equation 8 implies zero thermal noise. For a resistor, Equations 8 and 9 imply that the voltage noise spectrum is $S_U(f,T) = 4kTR$ and its Norton-equivalent, the thermal noise spectrum is $S_I(f,T) = 4kT/R$.

Let's assume that the impedance is a resistor $R_0$. Then the current driven resistor will produce the following voltage:

$$U(t) = R_0 I(t) + U_n(t) \quad , \tag{10}$$

where $U_n(t)$ is a Gaussian thermal noise with mean-square value $\int_{f_L}^{f_H} 4kTR_0 \, df$, where $f_L$ and $f_H$ are the low and high frequency cutoffs of the frequency band of observation, respectively.

## 3. The memristor equations are incomplete and unable to define a physical device

*3.1 The case of linear memristors*

There is a striking difference between Equations 6 and 10. Equation 6 with zero thermal noise for the memristor emulating a resistor is unphysical unless the memristor is an active device or, if it is passive, it must be at zero Kelvin absolute temperature to have zero noise. (Note, even the assumptions that an active impedance can have zero noise or that a physical body has zero absolute temperature are unphysical but let us disregard this problem).

To prove the above statement, let us suppose that the memristor is passive. Then in the arrangement shown in Figure 2, suppose that $Z_a$ and $Z_b$ are a memristor and a resistor, defined by Equations 6 and 10, respectively and both are at the same temperature $T$. As it is well known, it follows from equations 6 and 8-10, that in the frequency band $[f_L, f_H]$, the mean power flow between two identical resistors of $R_0$ at temperatures $T_1$ and $T_2$ is given as (e.g. [5-7]):

$$\langle P_{1 \to 2} \rangle_t = 4k(T_1 - T_2)(f_H - f_L)\frac{R_0^2}{4R_0^2} = k(T_1 - T_2)(f_H - f_L) \quad . \tag{11}$$

Then, if the memristor model provides a full description, there would be a non-zero mean power flow $\langle P_{R \to M}(T) \rangle_t$ from a passive resistor at temperature $T>0$ to a parallel memristor with zero thermal noise at the same temperature:

$$\langle P_{R \to M}(T) \rangle_t = kT(f_H - f_L) > 0 \quad . \tag{12}$$

However, both the memristor and the resistor are at the same temperature $T$ therefore a non-zero power flow violate the Second Law of Thermodynamics unless there is an external energy source that supplies this power. *The physical conclusion is* [4]*, that a memristor described by Equation 6 must be an active device.*

Alternatively, if the given memristor is actually a passive device then we must realize that the memristor equations are incomplete to address the statistical thermodynamics of the system. The statistical physics (Fluctuation-Dissipation Theorem) of the memristor is undefined and, without that, Equations 1-6 are not only incomplete but they are also incorrect if they are used to decide about the passivity/activity of the memristor.

*3.2 The case of nonlinear memristors*

The violation of the Second Law due to assuming passivity of the memristors and the validity of Equations 1-5 goes much beyond the case of linear memristors. Both from the above considerations and from [4] it is obvious, that assuming a thermal noise-free memristor, as Equations 1-6 do, will break the Second Law whenever the memristor function describes a dissipative device. Then, similarly to the linear case above, connecting a resistor parallel to the memristor will transfer non-zero net thermal noise power to the memristor from the resistor. As a consequence, the resistor is cooled and the memristor is heated.

Here, as the demonstration of infinite number of cases of activity, instead of investigating the general question of dissipativity of memristor functions, we point out that a smaller yet still infinitely large number of cases exists when the nonlinear memristor described by Equations 1-5 must be active: the case when an asymmetric memristor function shows rectifying characteristics, even at the slightest level, see Figure 3.

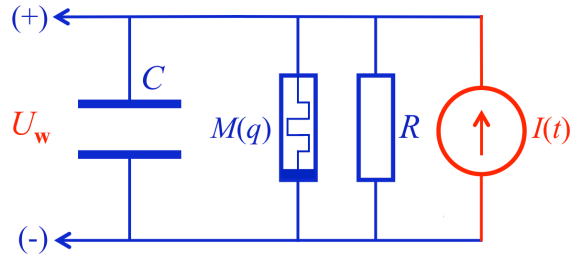

**Figure 3.** A DC voltage source by a rectifying-type memristor with the circuit model of the thermal noise current $I(t)$ of a shunt resistor.

If the thermal noise current flowing through the memristor and due to the proper asymmetry of the $\Phi$ and $M$ functions, the $U_w$ voltage on the capacitor has non-zero mean (DC) value, even if it has only an infinitesimally small DC component, a perpetual motion machine can be built. The same situation for an unphysical (that is, noise-free) diode was already pointed out by Brillouin [9].

We gave a proof of Brilluoin's paradox in [4] by showing that putting a large number $N$ of circuits shown in Figure 3 into a series circuitry results in an unbounded available power proportional to $N$ due to the DC components that are always positively correlated. In conclusion, any passive circuitry that can rectify thermal noise, even in the slightest way, in unphysical because it breaks the Second Law. Such circuit must contain an active element.

As an illustrative example, we created simple memristor functions that are able to rectify, see, Figure 4. It is obvious that these simple examples represent an infinite set of rectifying memristor functions

$$\Phi(q) = aq + bq^2 + cq^3 \tag{13}$$

if we choose the coefficients of *q* so that Equation 5 is satisfied an each of these coefficients are greater than zero.

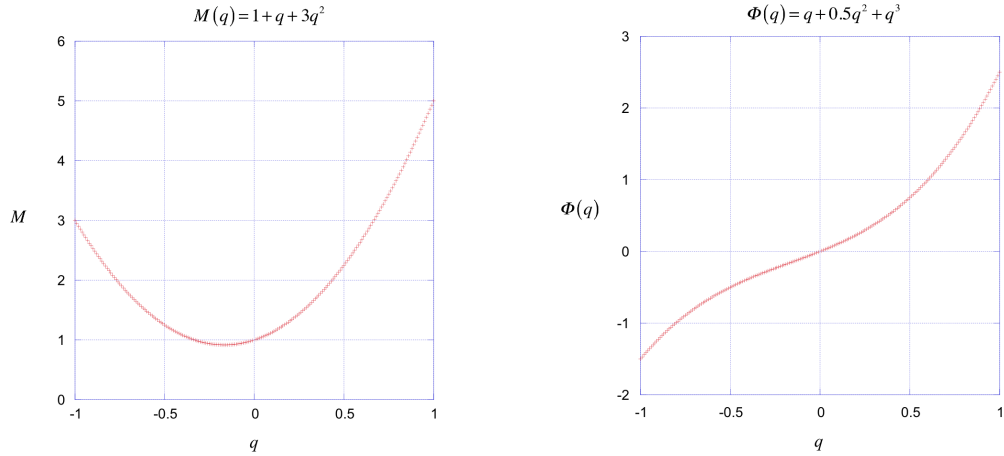

**Figure 4.** Memristor functions with rectifying characteristics. The asymmetry of *M* provides an asymmetry in $\Phi$ and a non-zero DC component in $U_w$ when driven by an AC current. In physical cases such *M* and $\Phi$ functions will be bounded at higher values however it is unnecessary to deal with this issue for thermal noise, which is the weakest signal in the system.

**4. Conclusion**

In their present noise-free form, the memristor equations (1-5) require the presence of an active device in infinitely many cases of memristor functions.

Chua's general proof that the memristor is passive based on Equations 1-5 is invalid because the memristor equations do not provide a complete description of the physics of memristors. A relevant statistical thermodynamics is needed for the memristors however the memristor equations do not contain sufficient physics to deduce that.

While the response functions of passive resistors, capacitors and inductors determine their Fluctuation-Dissipation Theorem (FDT), the memristor model lacks sufficient physics for a FDT. To create the noise theory (FDT) of a memristor, it is essential to know the internal material structure of the particular memristor thus its general formulation, like it is done with the three basic circuit elements, is impossible.

Finally, we note that, in a similar, nonlinear situation Nico van Kampen had already showed [10] that the vacuum diode equations lead to unphysical results if the thermal noise of the diode is not included. He deduced the correct theory of noise in freestanding diodes, which is a relevant illustration of the essentially missing component of memristor theories.

*A Post Script note*: It came to our attention during the publication of this paper that other Authors also objected the memristor claims by significant arguments which support our general observations about the non-physicality of the abstract mathematical memristor model. In 2012, Meuffels and Soni pointed out [11] that the HP-memristors must fail due to electrochemical arguments. In 2013, Di Ventra and Pershin discussed [12] that fluctuations would destroy the memory content of memristors. They attempted the impossible, to introduce a general Nyquist thermal noise theory for memristors. However they are using the Langevin approach to describe noise, which is forbidden in nonlinear elements with their own noise generation, see [13] a well-known general critique against such approach by van Kampen. This is why we believe (see above) that each memristor realization requires its own particular Fluctuation Dissipation Theorem. In 2015, Vongehr and Meng [14] criticized Chua's magnetic field arguments [1] and asserted "*The Missing Memristor has Not been Found*".